# Symmetry analysis of **k** = (1/3,1/3,0) three-sublattice ordering in $A_{14}B_{51}$ compounds

P. Fischer[1]*, V. Pomjakushin[1], L. Keller[1], A. Daoud[1], W. Sikora[2], A. Dommann[3,4], and F. Hulliger[3]

[1] Laboratory for Neutron Scattering, ETH Zurich & Paul Scherrer Institute, CH-5232 Villigen PSI, Switzerland

[2] Faculty of Physics and Applied Computer Science, AGH University of Science and Technology, PL-30-059 Krakow, Poland

[3] Laboratory for Solid State Physics, ETH, CH-8093 Zurich, Switzerland

[4] Present address: NTB, CH-9470 Buchs, Switzerland

**\* Corresponding author:**

Dr. Peter Fischer

Laboratory for Neutron Scattering, ETH Zurich & Paul Scherrer Institute, CH-5232 Villigen PSI, Switzerland

E-mail: Peter.Fischer@psi.ch

Tel.: +41564440306

Fax: +41564440750

**Abstract**

Bulk magnetic, X-ray and neutron diffraction measurements were performed on polycrystalline $Tb_{14}Ag_{51}$ in the temperature range from 1.5 K to room temperature. Its chemical $Gd_{14}Ag_{51}$ type structure corresponds to space group *P6/m*. Combined with group theoretical symmetry analysis, we show that the magnetic structure of this intermetallic compound is of a new **k** = (1/3,1/3,0) type with three magnetic Tb sublattices ordering simultaneously below $T_N$ = 27.5(5) K according to the combined irreducible representations $\tau_4$ and $\tau_6$. Here we present a general analysis of possible magnetic A ordering for **k** = (1/3,1/3,0) in $A_{14}B_{51}$ compounds.

## 1. Introduction

Intermetallic uranium and rare-earth $A_{14}B_{51}$ compounds with $Gd_{14}Ag_{51}$ structure [1] have interesting physical properties such as coexistence of antiferromagnetic order and heavy-fermion behavior in $Ce_{14}X_{51}$ (X = Au, Ag, Cu) [2] and in $U_{14}Au_{51}$. [3-5] This is related to the fact that there are three crystallographically distinct A sites in this structure.

Moreover, its particular hexagonal symmetry gives due to quasi triangular arrangement of magnetic ions rise to considerable geometric frustrations in the magnetic interactions. Thus from bulk physical measurements [2] it was concluded that generally not all Ce sublattices order magnetically in the antiferromagnetic $Ce_{14}X_{51}$ compounds with Néel temperatures $T_N \leq 3.2$ K. Also in $U_{14}Au_{51}$ with $T_N = 22$ K the (2e) sites do not exhibit magnetic ordering.

In 1997 the correct antiferromagnetic structure of this compound had been established by P. J. Brown et al. from single crystal neutron polarimetric and unpolarized neutron diffraction measurements [4] and had been later confirmed by µSR investigations performed by A. Schenck et al. . [5] The k = 0 magnetic structure was found to be non-collinear according to the Shubnikov group *P6/m'* with the U moments confined to the (a,b)-plane. The moments of U atoms in each of the two sets of six fold sites are arranged hexagonally with rotations of 60° between them and the two sets are rotated with respect to one another by 50°.

The mentioned interesting magnetic properties motivated us to start powder X-ray and neutron diffraction investigations as well as bulk magnetic measurements on the similar antiferromagnetic rare-earth compound $Tb_{14}Ag_{51}$ in order to determine to which extent magnetic ordering is different and in particular to investigate whether there may be also magnetic order on all rare-earth sites. Because of the larger magnetic Tb moments powder neutron diffraction is more appropriate to obtain first information on magnetic ordering than in $U_{14}Au_{51}$ [3] with rather small magnetic U moments and k = 0 magnetic structure. Indeed first neutron diffraction experiments performed by A. Dommann et al. [6] revealed **k** = (1/3,1/3,0) in case of $Tb_{14}Ag_{51}$. In the present work we obtained considerably improved neutron diffraction data and performed a careful analysis of both the chemical and magnetic structures of $Tb_{14}Ag_{51}$. In particular we shall prove that in $Tb_{14}Ag_{51}$ the magnetic ordering is of a new type in the important class of intermetallic $A_{14}B_{51}$ compounds with remarkable variation of physical properties. In contrast to the heavy fermion system $U_{14}Au_{51}$ [4], in $Tb_{14}Ag_{51}$ for the first time all three A sublattices are shown to order magnetically below $T_N$ = 27.5(5) K, cf. ref. . [7]

## 2. Group theory analysis of magnetic symmetry in case of three magnetic A sublattices of a $A_{14}B_{51}$ chemical structure for k-vector = (1/3,1/3,0)

The chemical structure of $A_{14}B_{51}$ compounds such as $Tb_{14}Ag_{51}$ corresponds to space group no. 175, *P6/m*. Without proper symmetry analysis it is rather hopeless to determine from powder neutron diffraction data the magnetic ordering in such a system with three magnetic A sublattices Tb1, Tb2 and Tb3 according to the sites (6k), (6j) and (2e), respectively (see Fig. 1). The chemical point symmetries of these sites are m, m and 6, respectively. As only one magnetic transition is observed as a function of temperature, we may assume that all three magnetic sublattices order simultaneously according to the same irreducible representations. As explained by Izyumov and Naish, [8] magnetic ordering may usually be attributed to a specific propagation vector **k** and to a single irreducible representation $\tau_j$ of the chemical structure. Moreover, considering the magnetic moments as complex axial vectors, the Fourier components of those in the nth chemical unit cell are related to ones in the 'zero' cell by equation (1).

(1)  $\Psi_{jn} = \Psi_{j0} e^{i\varphi(n)}$, where $\varphi(n) = 2\pi \mathbf{k}\mathbf{t}$, and **t**(n) is a translation vector from the 0-cell to the n-cell.

In general such relations with $e^{i2\pi \mathbf{k}\mathbf{t}}$ factors are not real quantities and thus cannot represent magnetic moments. To obtain real magnetic moments $\mathbf{S}(\mathbf{r}_j)$, one has to find suitable linear combinations, i.e. mixing coefficients of the basis vectors for **k** and -**k**, respectively, making use of equations (1) and (2). [9]

(2)  $\mathbf{S}(\mathbf{r}_j) = \sum_{\mu,\nu,\lambda} c_{\mu\nu\lambda} \Psi_{\mu\nu\lambda}(\mathbf{r}_j)$, where $\mu$ labels arms of the star of **k**, $\nu$ denotes the irreducible representations and $\lambda$ the dimension of the representation. { $\mathbf{r}_j$ } refers to an orbit. These coefficients may be used as order parameters in the analysis of magnetic phase transitions. [9]

From neutron diffraction results on $Tb_{14}Ag_{51}$ [6], we know **k** = (1/3,1/3,0). Based on the new Windows version of program MODY, [9] the corresponding basis vectors (magnetic modes $\psi$) are summarized in Table II for the three types of magnetic sites. The results were also checked by means of program BASIREP. [10]

The star of the vector **k** contains the two arms $\mathbf{k}_1 = \mathbf{k}$ and $\mathbf{k}_2 = (-1/3,2/3,0)$ which is equivalent to -**k**. The propagation vector **k** (small group G(**k**)) implies a symmetry reduction to space group $P\bar{6}$ with symmetry elements $1, 3^+, 3^-, m, \bar{6}^+$ and $\bar{6}^-$. For this case and space group $P6/m$, there are six one-dimensional irreducible representations $\tau_j$ (see Table II). The first two are real, and the other four are complex ($\tau_5 = \tau_3^*$, $\tau_6 = \tau_4^*$). Axial vector representations $\tau$ of the six magnetic moments with three components associated with both sites (6k) and (6j) decompose into two orbits of type $\tau = \tau_1+2\tau_2+\tau_3+2\tau_4+\tau_5+2\tau_6$. On the other hand for sites (2e) only one orbit exits: $\tau = \tau_1+\tau_2+\tau_3+\tau_4+\tau_5+\tau_6$. Thus in case of simultaneous magnetic ordering of all three sites, all $\tau_j$ are possible for each site. Furthermore it should be noted that for the present space group (for irreducible representations $\tau_3$, $\tau_4$, $\tau_5$ and $\tau_6$ with Herring coefficient 0), one has to combine in the sense of a direct sum $\tau_3 \oplus \tau_5$ magnetic modes (basis vectors) for +**k** from $\tau_3$ with those for –**k** from $\tau_5$ and vice-versa (similar for $\tau_4 \oplus \tau_6$).

With respect to possible magnetic configurations resulting from Table II, we discuss here in more detail those for the ‚complementary' (z- resp. (x,y)-components of the magnetic moments) irreducible representations $\tau_1$ and $\tau_4 \oplus \tau_6$. The latter will be shown to hold for $Tb_{14}Ag_{51}$. For further details of the calculations and other cases we refer to the appendix. It should be noted that the two orbits in case of sites (6k) and (6j) are with respect to only symmetry considerations independent. This is indicated by the different phases. Also the moment magnitudes could differ, but we consider this in view of probable exchange and dipolar interactions as less realistic.

For sites (6k) and irreducible representation $\tau_1$ one obtains from Table II (see appendix) the solution:

(6k1)    $S_{1n} = S_{2'n} = S_{3'n} = S_1 e_z \cos(\varphi(n)+\alpha_1)$; $S_{4'n} = S_{5'n} = S_{6n} = S_1 e_z \cos(\varphi(n)+\beta_1)$,

where $e_z$ = unit vector along the **c**-direction.

For one magnetic site such as (6k) normally the phase angle $\alpha_1$ can be set to zero, as the neutron intensities depend only on the relative phase $\beta_1$. Thus the configuration (6k1) represents an amplitude modulated antiferromagnetic structure with the magnetic moments oriented parallel to the **c**-axis (z-components for $\alpha_1 = 0$: $+S_1, -S_1/2, -S_1/2$) along the basic translations **a** and **b** of the chemical unit cell. This yields a nine times larger magnetic unit cell, compared to the chemical unit cell.

For sites (6j) and irreducible representation $\tau_1$ one obtains a similar solution:

(6j1)  $\mathbf{S}_{1n} = \mathbf{S}_{2'n} = \mathbf{S}_{3'n} = S_2\mathbf{e}_z\cos(\varphi(n)+\alpha_2); \mathbf{S}_{4n} = \mathbf{S}_{5'n} = \mathbf{S}_{6'n} = S_2\mathbf{e}_z\cos(\varphi(n)+\beta_2).$

Concerning the third kind of sites (2e) results for $\tau_1$:

(2e1)  $\mathbf{S}_{1n} = S_3\mathbf{e}_z\cos(\varphi(n)+\alpha_3) = \mathbf{S}_{2n}.$

Whereas $\tau_1$ confines the magnetic moments to be oriented parallel to the c-axis, they are restricted to the basal plane in case of $\tau_4 \oplus \tau_6$:

One obtains for sites (6k) three basic modulated solutions, using +k from $\tau_4$:

(6k46a)
$$\mathbf{S}_{1n} = S_1\mathbf{e}_x\cos(\varphi(n)+\alpha_{1a}), \qquad \mathbf{S}_{4'n} = S_1\mathbf{e}_x\cos(\varphi(n)+\beta_{1a}),$$
$$\mathbf{S}_{2'n} = -S_1\mathbf{e}_y\cos(\varphi(n)+\pi/3+\alpha_{1a}), \qquad \mathbf{S}_{5'n} = -S_1\mathbf{e}_y\cos(\varphi(n)+\pi/3+\beta_{1a}),$$
$$\mathbf{S}_{3'n} = S_1(\mathbf{e}_x+\mathbf{e}_y)\cos(\varphi(n)-\pi/3+\alpha_{1a}), \qquad \mathbf{S}_{6n} = S_1(\mathbf{e}_x+\mathbf{e}_y)\cos(\varphi(n)-\pi/3+\beta_{1a}),$$

where in each orbit the magnetic moments are oriented parallel.

(6k46b)
$$\mathbf{S}_{1n} = S_1\mathbf{e}_y\cos(\varphi(n)+\alpha_{1b}), \qquad \mathbf{S}_{4'n} = S_1\mathbf{e}_y\cos(\varphi(n)+\beta_{1b}),$$
$$\mathbf{S}_{2'n} = S_1(\mathbf{e}_x+\mathbf{e}_y)\cos(\varphi(n)+\pi/3+\alpha_{1b}), \qquad \mathbf{S}_{5'n} = S_1(\mathbf{e}_x+\mathbf{e}_y)\cos(\varphi(n)+\pi/3+\beta_{1b}),$$
$$\mathbf{S}_{3'n} = -S_1\mathbf{e}_x\cos(\varphi(n)-\pi/3+\alpha_{1b}), \qquad \mathbf{S}_{6n} = -S_1\mathbf{e}_x\cos(\varphi(n)-\pi/3+\beta_{1b}),$$

(6k46c)
$$\mathbf{S}_{1n} = S_1(\mathbf{e}_x+\mathbf{e}_y)\cos(\varphi(n)+\alpha_{1c}), \qquad \mathbf{S}_{4'n} = S_1(\mathbf{e}_x+\mathbf{e}_y)\cos(\varphi(n)+\beta_{1c}),$$
$$\mathbf{S}_{2'n} = S_1\mathbf{e}_x\cos(\varphi(n)+\pi/3+\alpha_{1c}), \qquad \mathbf{S}_{5'n} = S_1\mathbf{e}_x\cos(\varphi(n)+\pi/3+\beta_{1c}),$$
$$\mathbf{S}_{3'n} = S_1\mathbf{e}_y\cos(\varphi(n)-\pi/3+\alpha_{1c}), \qquad \mathbf{S}_{6n} = S_1\mathbf{e}_y\cos(\varphi(n)-\pi/3+\beta_{1c}).$$

By further combination such as $[(6k46a)+2(6k46b)]/\sqrt{3}$ results for the first orbit:

(6k46d)
$$\mathbf{S}_{1n} = S_1(1/\sqrt{3})(\mathbf{e}_x+2\mathbf{e}_y)\cos(\varphi(n)+\alpha_{1d}),$$
$$\mathbf{S}_{2'n} = S_1(1/\sqrt{3})(2\mathbf{e}_x+\mathbf{e}_y)\cos(\varphi(n)+\pi/3+\alpha_{1d}),$$
$$\mathbf{S}_{3'n} = S_1(1/\sqrt{3})(-\mathbf{e}_x+\mathbf{e}_y)\cos(\varphi(n)-\pi/3+\alpha_{1d}).$$

Similar relations hold for the second orbit with 4'n, 5'n and 6n instead of 1n, 2'n and 3'n as well as $\beta$ instead of $\alpha$.

As the magnetic moment directions in (6k46a) and (6k46d) are perpendicular, we may form with $\alpha_{1a} = \alpha_1$, $\beta_{1a} = \beta_1$, $\alpha_{1d} = \alpha_1-\pi/2$ and $\beta_{1d} = \beta_1-\pi/2$ also cycloidal spiral solutions with constant magnitude of the magnetic Tb moments such as:

$$\mathbf{S}_{1n} = S_1[\mathbf{e}_x\cos(\varphi(n)+\alpha_1)+(1/\sqrt{3})(\mathbf{e}_x+2\mathbf{e}_y)\sin(\varphi(n)+\alpha_1)],$$

$$\mathbf{S}_{2'n} = -S_1[\mathbf{e}_y\cos(\varphi(n)+\pi/3+\alpha_1)-(1/\sqrt{3})(2\mathbf{e}_x+\mathbf{e}_y)\sin(\varphi(n)+\pi/3+\alpha_1)],$$

$$\mathbf{S}_{3'n} = S_1[(\mathbf{e}_x+\mathbf{e}_y)\cos(\varphi(n)-\pi/3+\alpha_1)-(1/\sqrt{3})(\mathbf{e}_x-\mathbf{e}_y)\sin(\varphi(n)-\pi/3+\alpha_1)],$$

(6k46e)

$$\mathbf{S}_{4'n} = S_1[\mathbf{e}_x\cos(\varphi(n)+\beta_1)+(1/\sqrt{3})(\mathbf{e}_x+2\mathbf{e}_y)\sin(\varphi(n)+\beta_1)],$$

$$\mathbf{S}_{5'n} = -S_1[\mathbf{e}_y\cos(\varphi(n)+\pi/3+\beta_1)-(1/\sqrt{3})(2\mathbf{e}_x+\mathbf{e}_y)\sin(\varphi(n)+\pi/3+\beta_1)],$$

$$\mathbf{S}_{6n} = S_1[(\mathbf{e}_x+\mathbf{e}_y)\cos(\varphi(n)-\pi/3+\beta_1)-(1/\sqrt{3})(\mathbf{e}_x-\mathbf{e}_y)\sin(\varphi(n)-\pi/3+\beta_1)].$$

They are particularly reasonable with respect to the fact that the 4f electrons of heavy rare earths such as $Tb^{3+}$ appear to be generally rather well localized. Appropriate choice of free phases implies that other solutions resulting from (6k46a) to (6k46c) become equivalent to (6k46e).

General combinations u(6k46a)+v(6k46b) and u(6k46c)+v(6k46a) (needed for FULLPROF [10]) with respect to hexagonal basic translations result for the two orbits with $\alpha_{1a} = \alpha_{1b} = \alpha_1$ and $\beta_1 = \beta_{1c}+\pi/3 = \beta_{1a}-2\pi/3$ in the following transformation matrices of magnetic moments:

(6k46f)

| x,y,z: (u,v,w);phase 0, | -x,-y,z: (u,v,w);0, |
| -y,x-y,z: (v,-u+v,w);+π/3, | y,-x+y,z: (v,-u+v,w);+π/3, |
| -x+y,-x,z: (u-v,u,w);-π/3, | x-y,x,z: (u-v,u,w);-π/3. |

It may be used both for modulated and constant magnitude configurations.

Starting from the +**k** magnetic modes from $\tau_6$ yields:

9$$\mathbf{S}_{1n} = S_1\mathbf{e}_x\cos(\varphi(n)+\alpha_{1a}'), \qquad \mathbf{S}_{4'n} = S_1\mathbf{e}_x\cos(\varphi(n)+\beta_{1a}'),$$

$$(6k46a')\ \mathbf{S}_{2'n} = -S_1\mathbf{e}_y\cos(\varphi(n)-\pi/3+\alpha_{1a}'), \qquad \mathbf{S}_{5'n} = -S_1\mathbf{e}_y\cos(\varphi(n)-\pi/3+\beta_{1a}'),$$

$$\mathbf{S}_{3'n} = S_1(\mathbf{e}_x+\mathbf{e}_y)\cos(\varphi(n)+\pi/3+\alpha_{1a}'), \ \mathbf{S}_{6n} = S_1(\mathbf{e}_x+\mathbf{e}_y)\cos(\varphi(n)+\pi/3+\beta_{1a}'),$$

$$\mathbf{S}_{1n} = S_1\mathbf{e}_y\cos(\varphi(n)+\alpha_{1b}'), \qquad \mathbf{S}_{4'n} = S_1\mathbf{e}_y\cos(\varphi(n)+\beta_{1b}'),$$

$$(6k46b')\ \mathbf{S}_{2'n} = S_1(\mathbf{e}_x+\mathbf{e}_y)\cos(\varphi(n)-\pi/3+\alpha_{1b}'),\ \mathbf{S}_{5'n} = S_1(\mathbf{e}_x+\mathbf{e}_y)\cos(\varphi(n)-\pi/3+\beta_{1b}'),$$

$$\mathbf{S}_{3'n} = -S_1\mathbf{e}_x\cos(\varphi(n)+\pi/3+\alpha_{1b}'), \qquad \mathbf{S}_{6n} = -S_1\mathbf{e}_x\cos(\varphi(n)+\pi/3+\beta_{1b}'),$$

$$\mathbf{S}_{1n} = S_1(\mathbf{e}_x+\mathbf{e}_y)\cos(\varphi(n)+\alpha_{1c}'), \qquad \mathbf{S}_{4'n} = S_1(\mathbf{e}_x+\mathbf{e}_y)\cos(\varphi(n)+\beta_{1c}'),$$

$$(6k46c')\ \mathbf{S}_{2'n} = S_1\mathbf{e}_x\cos(\varphi(n)-\pi/3+\alpha_{1c}'), \qquad \mathbf{S}_{5'n} = S_1\mathbf{e}_x\cos(\varphi(n)-\pi/3+\beta_{1c}'),$$

$$\mathbf{S}_{3'n} = S_1\mathbf{e}_y\cos(\varphi(n)+\pi/3+\alpha_{1c}'), \qquad \mathbf{S}_{6n} = S_1\mathbf{e}_y\cos(\varphi(n)+\pi/3+\beta_{1c}').$$

In each orbit the magnetic moments form a triangle with rotation sense to the right.

By further combination such as $[(6k46a')+2(6k46b')]/\sqrt{3}$ results:

$$\mathbf{S}_{1n} = S_1(1/\sqrt{3})(\mathbf{e}_x+2\mathbf{e}_y)\cos(\varphi(n)+\alpha_{1d}'),$$

$$(6k46d')\ \mathbf{S}_{2'n} = S_1(1/\sqrt{3})(2\mathbf{e}_x+\mathbf{e}_y)\cos(\varphi(n)-\pi/3+\alpha_{1d}'),$$

$$\mathbf{S}_{3'n} = S_1(1/\sqrt{3})(-\mathbf{e}_x+\mathbf{e}_y)\cos(\varphi(n)+\pi/3+\alpha_{1d}').$$

Similar relations hold for the second orbit.

As the magnetic moment directions in (6k46a') and (6k46d') are perpendicular, we may form further cycloidal spiral solutions with constant magnitude of the magnetic Tb moments such as:

$$\mathbf{S}_{1n} = S_1[\mathbf{e}_x\cos(\varphi(n)+\alpha_1)+(1/\sqrt{3})(\mathbf{e}_x+2\mathbf{e}_y)\sin(\varphi(n)+\alpha_1)],$$

$$\mathbf{S}_{2'n} = -S_1[\mathbf{e}_y\cos(\varphi(n)-\pi/3+\alpha_1)-(1/\sqrt{3})(2\mathbf{e}_x+\mathbf{e}_y)\sin(\varphi(n)-\pi/3+\alpha_1)],$$

$$\mathbf{S}_{3'n} = S_1[(\mathbf{e}_x+\mathbf{e}_y)\cos(\varphi(n)+\pi/3+\alpha_1)-(1/\sqrt{3})(\mathbf{e}_x-\mathbf{e}_y)\sin(\varphi(n)+\pi/3+\alpha_1)],$$

(6k46e')

$$\mathbf{S}_{4'n} = S_1[\mathbf{e}_x\cos(\varphi(n)+\beta_1)+(1/\sqrt{3})(\mathbf{e}_x+2\mathbf{e}_y)\sin(\varphi(n)+\beta_1)],$$

$$\mathbf{S}_{5'n} = -S_1[\mathbf{e}_y\cos(\varphi(n)-\pi/3+\beta_1)-(1/\sqrt{3})(2\mathbf{e}_x+\mathbf{e}_y)\sin(\varphi(n)-\pi/3+\beta_1)],$$

$$\mathbf{S}_{6n} = S_1[(\mathbf{e}_x+\mathbf{e}_y)\cos(\varphi(n)+\pi/3+\beta_1)-(1/\sqrt{3})(\mathbf{e}_x-\mathbf{e}_y)\sin(\varphi(n)+\pi/3+\beta_1)].$$

General combinations u(6k46a')+v(6k46b') and u(6k46c')+v(6k46a') result for the two orbits with $\alpha_{1a}' = \alpha_{1b}' = \alpha_1$ and $\beta_1 = \beta_{1c}'-\pi/3 = \beta_{1a}'-4\pi/3$ in the following transformation matrices of magnetic moments:

(6k46f')
| x,y,z: (u,v,w);0, | -x,-y,z: (u,v,w);0, |
| -y,x-y,z: (v,-u+v,w);-π/3, | y,-x+y,z: (v,-u+v,w);-π/3, |
| -x+y,-x,z: (u-v,u,w);+π/3, | x-y,x,z: (u-v,u,w);+π/3. |

Starting from +**k** from $\tau_4$, one obtains for sites (6j) similar magnetic configurations as in case of sites (6k):

(6j46a)
$$\mathbf{S}_{1n} = S_2\mathbf{e}_x\cos(\varphi(n)+\alpha_{2a}), \qquad \mathbf{S}_{4n} = S_2\mathbf{e}_x\cos(\varphi(n)+\beta_{2a}),$$
$$\mathbf{S}_{2'n} = -S_2\mathbf{e}_y\cos(\varphi(n)+\pi/3+\alpha_{2a}), \qquad \mathbf{S}_{5'n} = -S_2\mathbf{e}_y\cos(\varphi(n)+\pi/3+\beta_{2a}),$$
$$\mathbf{S}_{3'n} = S_2(\mathbf{e}_x+\mathbf{e}_y)\cos(\varphi(n)-\pi/3+\alpha_{2a}), \quad \mathbf{S}_{6'n} = S_2(\mathbf{e}_x+\mathbf{e}_y)\cos(\varphi(n)-\pi/3+\beta_{2a}),$$

where the magnetic moments in each orbit are oriented parallel.

(6j46b)
$$\mathbf{S}_{1n} = S_2\mathbf{e}_y\cos(\varphi(n)+\alpha_{2b}), \qquad \mathbf{S}_{4n} = S_2\mathbf{e}_y\cos(\varphi(n)+\beta_{2b}),$$
$$\mathbf{S}_{2'n} = S_2(\mathbf{e}_x+\mathbf{e}_y)\cos(\varphi(n)+\pi/3+\alpha_{2b}), \quad \mathbf{S}_{5'n} = S_2(\mathbf{e}_x+\mathbf{e}_y)\cos(\varphi(n)+\pi/3+\beta_{2b}),$$
$$\mathbf{S}_{3'n} = -S_2\mathbf{e}_x\cos(\varphi(n)-\pi/3+\alpha_{2b}), \qquad \mathbf{S}_{6'n} = -S_2\mathbf{e}_x\cos(\varphi(n)-\pi/3+\beta_{2b}),$$

$$\mathbf{S}_{1n} = S_2(\mathbf{e}_x+\mathbf{e}_y)\cos(\varphi(n)+\alpha_{2c}), \qquad \mathbf{S}_{4n} = S_2(\mathbf{e}_x+\mathbf{e}_y)\cos(\varphi(n)+\beta_{2c}),$$

(6j46c) $\quad \mathbf{S}_{2'n} = S_2\mathbf{e}_x\cos(\varphi(n)+\pi/3+\alpha_{2c}), \qquad \mathbf{S}_{5'n} = S_2\mathbf{e}_x\cos(\varphi(n)+\pi/3+\beta_{2c}),$

$$\mathbf{S}_{3'n} = S_2\mathbf{e}_y\cos(\varphi(n)-\pi/3+\alpha_{2c}), \qquad \mathbf{S}_{6'n} = S_2\mathbf{e}_y\cos(\varphi(n)-\pi/3+\beta_{2c}).$$

$$\mathbf{S}_{1n} = S_2(1/\sqrt{3})(\mathbf{e}_x+2\mathbf{e}_y)\cos(\varphi(n)+\alpha_{2d}),$$

(6j46d) $\quad \mathbf{S}_{2'n} = S_2(1/\sqrt{3})(2\mathbf{e}_x+\mathbf{e}_y)\cos(\varphi(n)+\pi/3+\alpha_{2d}),$

$$\mathbf{S}_{3'n} = S_2(1/\sqrt{3})(-\mathbf{e}_x+\mathbf{e}_y)\cos(\varphi(n)-\pi/3+\alpha_{2d}).$$

Similar relations hold for the second orbit, yielding also:

$$\mathbf{S}_{1n} = S_2[\mathbf{e}_x\cos(\varphi(n)+\alpha_2)+(1/\sqrt{3})(\mathbf{e}_x+2\mathbf{e}_y)\sin(\varphi(n)+\alpha_2)],$$

$$\mathbf{S}_{2'n} = -S_2[\mathbf{e}_y\cos(\varphi(n)+\pi/3+\alpha_2)-(1/\sqrt{3})(2\mathbf{e}_x+\mathbf{e}_y)\sin(\varphi(n)+\pi/3+\alpha_2)],$$

$$\mathbf{S}_{3'n} = S_2[(\mathbf{e}_x+\mathbf{e}_y)\cos(\varphi(n)-\pi/3+\alpha_2)-(1/\sqrt{3})(\mathbf{e}_x-\mathbf{e}_y)\sin(\varphi(n)-\pi/3+\alpha_2)],$$

(6j46e)

$$\mathbf{S}_{4n} = S_2[\mathbf{e}_x\cos(\varphi(n)+\beta_2)+(1/\sqrt{3})(\mathbf{e}_x+2\mathbf{e}_y)\sin(\varphi(n)+\beta_2)],$$

$$\mathbf{S}_{5'n} = -S_2[\mathbf{e}_y\cos(\varphi(n)+\pi/3+\beta_2)-(1/\sqrt{3})(2\mathbf{e}_x+\mathbf{e}_y)\sin(\varphi(n)+\pi/3+\beta_2)],$$

$$\mathbf{S}_{6'n} = S_2[(\mathbf{e}_x+\mathbf{e}_y)\cos(\varphi(n)-\pi/3+\beta_2)-(1/\sqrt{3})(\mathbf{e}_x-\mathbf{e}_y)\sin(\varphi(n)-\pi/3+\beta_2)].$$

(6j46f)
$\quad$ x,y,z: (u,v,w); phase 0, $\qquad$ -x,-y,z: (u,v,w); 0,
$\quad$ -y,x-y,z: (v,-u+v,w); +π/3, $\qquad$ y,-x+y,z: (v,-u+v,w); +π/3,
$\quad$ -x+y,-x,z: (u-v,u,w); -π/3, $\qquad$ x-y,x,z: (u-v,u,w); -π/3.

The +**k** magnetic modes from $\tau_6$ yield:

$$\mathbf{S}_{1n} = S_2\mathbf{e}_x\cos(\varphi(n)+\alpha_{2a}'), \qquad \mathbf{S}_{4n} = S_2\mathbf{e}_x\cos(\varphi(n)+\beta_{2a}'),$$

(6j46a') $\quad \mathbf{S}_{2'n} = -S_2\mathbf{e}_y\cos(\varphi(n)-\pi/3+\alpha_{2a}'), \qquad \mathbf{S}_{5'n} = -S_2\mathbf{e}_y\cos(\varphi(n)-\pi/3+\beta_{2a}'),$

$$\mathbf{S}_{3'n} = S_2(\mathbf{e}_x+\mathbf{e}_y)\cos(\varphi(n)+\pi/3+\alpha_{2a}'), \; \mathbf{S}_{6'n} = S_2(\mathbf{e}_x+\mathbf{e}_y)\cos(\varphi(n)+\pi/3+\beta_{2a}'),$$

$$\mathbf{S}_{1n} = S_2\mathbf{e}_y\cos(\varphi(n)+\alpha_{2b}'), \qquad \mathbf{S}_{4n} = S_2\mathbf{e}_y\cos(\varphi(n)+\beta_{2b}'),$$

(6j46b') $\mathbf{S}_{2'n} = S_2(\mathbf{e}_x+\mathbf{e}_y)\cos(\varphi(n)-\pi/3+\alpha_{2b}'),\ \mathbf{S}_{5'n} = S_2(\mathbf{e}_x+\mathbf{e}_y)\cos(\varphi(n)-\pi/3+\beta_{2b}'),$

$$\mathbf{S}_{3'n} = -S_2\mathbf{e}_x\cos(\varphi(n)+\pi/3+\alpha_{2b}'), \qquad \mathbf{S}_{6'n} = -S_2\mathbf{e}_x\cos(\varphi(n)+\pi/3+\beta_{2b}'),$$

$$\mathbf{S}_{1n} = S_2(\mathbf{e}_x+\mathbf{e}_y)\cos(\varphi(n)+\alpha_{2c}'), \qquad \mathbf{S}_{4n} = S_2(\mathbf{e}_x+\mathbf{e}_y)\cos(\varphi(n)+\beta_{2c}'),$$

(6j46c') $\mathbf{S}_{2'n} = S_2\mathbf{e}_x\cos(\varphi(n)-\pi/3+\alpha_{2c}'), \qquad \mathbf{S}_{5'n} = S_2\mathbf{e}_x\cos(\varphi(n)-\pi/3+\beta_{2c}'),$

$$\mathbf{S}_{3'n} = S_2\mathbf{e}_y\cos(\varphi(n)+\pi/3+\alpha_{2c}'), \qquad \mathbf{S}_{6'n} = S_2\mathbf{e}_y\cos(\varphi(n)+\pi/3+\beta_{2c}').$$

$$\mathbf{S}_{1n} = S_2(1/\sqrt{3})(\mathbf{e}_x+2\mathbf{e}_y)\cos(\varphi(n)+\alpha_{2d}),$$

(6j46d') $\mathbf{S}_{2'n} = S_2(1/\sqrt{3})(2\mathbf{e}_x+\mathbf{e}_y)\cos(\varphi(n)-\pi/3+\alpha_{2d}),$

$$\mathbf{S}_{3'n} = S_2(1/\sqrt{3})(-\mathbf{e}_x+\mathbf{e}_y)\cos(\varphi(n)+\pi/3+\alpha_{2d}).$$

Similar relations hold for the second orbit, leading also to:

$$\mathbf{S}_{1n} = S_2[\mathbf{e}_x\cos(\varphi(n)+\alpha_2)+(1/\sqrt{3})(\mathbf{e}_x+2\mathbf{e}_y)\sin(\varphi(n)+\alpha_2)],$$

$$\mathbf{S}_{2'n} = -S_2[\mathbf{e}_y\cos(\varphi(n)-\pi/3+\alpha_2)-(1/\sqrt{3})(2\mathbf{e}_x+\mathbf{e}_y)\sin(\varphi(n)-\pi/3+\alpha_2)],$$

$$\mathbf{S}_{3'n} = S_2[(\mathbf{e}_x+\mathbf{e}_y)\cos(\varphi(n)+\pi/3+\alpha_2)-(1/\sqrt{3})(\mathbf{e}_x-\mathbf{e}_y)\sin(\varphi(n)+\pi/3+\alpha_2)],$$

(6j46e')

$$\mathbf{S}_{4n} = S_2[\mathbf{e}_x\cos(\varphi(n)+\beta_2)+(1/\sqrt{3})(\mathbf{e}_x+2\mathbf{e}_y)\sin(\varphi(n)+\beta_2)],$$

$$\mathbf{S}_{5'n} = -S_2[\mathbf{e}_y\cos(\varphi(n)-\pi/3+\beta_2)-(1/\sqrt{3})(2\mathbf{e}_x+\mathbf{e}_y)\sin(\varphi(n)-\pi/3+\beta_2)],$$

$$\mathbf{S}_{6'n} = S_2[(\mathbf{e}_x+\mathbf{e}_y)\cos(\varphi(n)+\pi/3+\beta_2)-(1/\sqrt{3})(\mathbf{e}_x-\mathbf{e}_y)\sin(\varphi(n)+\pi/3+\beta_2)].$$

$$\begin{array}{ll} x,y,z: (u,v,w);0, & -x,-y,z: (u,v,w);0, \end{array}$$

(6j46f') $-y,x-y,z: (v,-u+v,w);-\pi/3, \qquad y,-x+y,z: (v,-u+v,w);-\pi/3,$

$$-x+y,-x,z: (u-v,u,w);+\pi/3, \qquad x-y,x,z: (u-v,u,w);+\pi/3.$$

Concerning the third kind of sites (2e) results for $\tau_4 \oplus \tau_6$:

(2e46)  $\mathbf{S}_{1n} = S_3[\mathbf{e}_x\cos(\varphi(n)+\alpha_3)+(1/\sqrt{3})(\mathbf{e}_x+2\mathbf{e}_y)\sin(\varphi(n)+\alpha_3)] = \mathbf{S}_{2n}$.

using +**k** from $\tau_4$ and

(2e46')  $\mathbf{S}_{1n} = S_3\mathbf{e}_x\cos(\varphi(n)+\alpha_3) = \mathbf{S}_{2n}$ for +**k** from both $\tau_4$ and $\tau_6$.

As the cosine and sine terms correspond to perpendicular directions, one thus again obtains the possibility of a triangular moment configuration in the basal plane with constant magnitude of the magnetic moments.

The discussed models show that the particular **k**-vector implies in a natural way a decomposition into three magnetic sublattices A, B and C, either by 120° rotations in case of planar models or by characteristic phase shifts for amplitude modulation. Thus one may choose a hexagonal H cell as smallest magnetic unit cell, but for convenience we keep the compared to the chemical unit cell along the **a**- and **b**-directions tripled magnetic unit cell. As in this cell translations 0,0,0; 2/3,1/3,0 and 1/3,2/3,0 belong to the same sublattice, one obtains the rule –h + k = 3n, n = integer for allowed magnetic peaks.

Finally we would like to mention that a similar symmetry analysis for k = 0 shows that the $U_{14}Au_{51}$ magnetic structure [4] corresponds to the real irreducible representation $\tau_2$.

**Appendix:** Examples of mixing coefficients and further possible magnetic configurations.

1) Sites (6k):

For sites (6k) and irreducible representation $\tau_1$ appropriate mixing coefficients are $c_1 = 0.5 S_1(m_1, -a^* m_1^*)$ and $c_2 = 0.5 S_1(-n_1 a^*, n_1^*)$, $m_1 = e^{i\alpha_1}$, $n_1 = e^{i\beta_1}$ for the two independent orbits. The first and second terms in the coefficient parentheses correspond to **k** and -**k**, to be multiplied with $e^{i\varphi(n)}$ and $e^{-i\varphi(n)}$, respectively, yielding

$\mathbf{S}_{1n} = (S_1/2)(e^{i(2\pi \mathbf{k}t+\alpha_1)}\psi_{6k}(\tau_1,\mathbf{k}) + e^{-i(2\pi \mathbf{k}t+\alpha_1)}\psi_{6k}(\tau_1,-\mathbf{k})) = S_1 \mathbf{e}_z \cos(\varphi(n)+\alpha_1)$ etc., where $\psi$ denotes a basic magnetic mode function, see Table II.

In case of $\tau_2$ one obtains in a similar straightforward way (e.g. $c_1 = 0.5 S_1(m_{1a}, -a^* m_{1a}^*; 0, -a^* m_{1a}^*)$, $c_2 = 0.5 S_1(0, a^* n_{1a}^*; n_{1a}, 0)$, $m_{1a} = e^{i\alpha_{1a}}$, $n_{1a} = e^{i\beta_{1a}}$) the solutions:

$$\mathbf{S}_{1n} = S_1 \mathbf{e}_x \cos(\varphi(n)+\alpha_{1a}), \qquad \mathbf{S}_{4'n} = S_1 \mathbf{e}_x \cos(\varphi(n)+\beta_{1a}),$$

(6k2a) $\quad \mathbf{S}_{2'n} = S_1 \mathbf{e}_y \cos(\varphi(n)+\alpha_{1a}), \qquad \mathbf{S}_{5'n} = S_1 \mathbf{e}_y \cos(\varphi(n)+\beta_{1a}),$

$$\mathbf{S}_{3'n} = -S_1(\mathbf{e}_x+\mathbf{e}_y)\cos(\varphi(n)+\alpha_{1a}), \qquad \mathbf{S}_{6n} = -S_1(\mathbf{e}_x+\mathbf{e}_y)\cos(\varphi(n)+\beta_{1a}),$$

where the magnetic moments in each group form a regular triangle with left-rotation by 120°.

$$\mathbf{S}_{1n} = S_1 \mathbf{e}_y \cos(\varphi(n)+\alpha_{1b}), \qquad \mathbf{S}_{4'n} = S_1 \mathbf{e}_y \cos(\varphi(n)+\beta_{1b}),$$

(6k2b) $\quad \mathbf{S}_{2'n} = -S_1(\mathbf{e}_x+\mathbf{e}_y)\cos(\varphi(n)+\alpha_{1b}), \qquad \mathbf{S}_{5'n} = -S_1(\mathbf{e}_x+\mathbf{e}_y)\cos(\varphi(n)+\beta_{1b}),$

$$\mathbf{S}_{3'n} = S_1 \mathbf{e}_x \cos(\varphi(n)+\alpha_{1b}), \qquad \mathbf{S}_{6n} = S_1 \mathbf{e}_x \cos(\varphi(n)+\beta_{1b}),$$

$$S_{1n} = S_1(e_x+e_y)\cos(\varphi(n)+\alpha_{1c}), \qquad S_{4'n} = S_1(e_x+e_y)\cos(\varphi(n)+\beta_{1c}),$$

(6k2c) $$S_{2'n} = -S_1 e_x \cos(\varphi(n)+\alpha_{1c}), \qquad S_{5'n} = -S_1 e_x \cos(\varphi(n)+\beta_{1c}),$$

$$S_{3'n} = -S_1 e_y \cos(\varphi(n)+\alpha_{1c}), \qquad S_{6n} = -S_1 e_y \cos(\varphi(n)+\beta_{1c}).$$

Similar to (6k46e) etc. we may thus form also constant magnetic moment solutions in the (a,b)-plane and construct the transformation matrices for magnetic moments:

x,y,z: (u,v,w);0,   -x,-y,z: (u,v,w);0,

(6k2f)  -y,x-y,z: (-v,u-v,w);0,   y,-x+y,z: (-v,u-v,w);0,

-x+y,-x,z: (-u+v,-u,w);0,   x-y,x,z: (-u+v,-u,w);0.

In case of $\tau_3 \oplus \tau_5$ one gets with +**k** from $\tau_3$:

$$S_{1n} = S_1 e_z \cos(\varphi(n)+\alpha_1), \qquad S_{4'n} = S_1 e_z \cos(\varphi(n)+\beta_1),$$

(6k35) $$S_{2'n} = -S_1 e_z \cos(\varphi(n)+\pi/3+\alpha_1), \qquad S_{5'n} = -S_1 e_z \cos(\varphi(n)+\pi/3+\beta_1),$$

$$S_{3'n} = -S_1 e_z \cos(\varphi(n)-\pi/3+\alpha_1), \qquad S_{6n} = -S_1 e_z \cos(\varphi(n)-\pi/3+\beta_1) \text{ and}$$

with +**k** from $\tau_5$:

$$S_{1n} = S_1 e_z \cos(\varphi(n)+\alpha_1'), \qquad S_{4'n} = S_1 e_x \cos(\varphi(n)+\beta_1'),$$

(6k35') $$S_{2'n} = -S_1 e_z \cos(\varphi(n)-\pi/3+\alpha_1'), \qquad S_{5'n} = S_1 e_y \cos(\varphi(n)-\pi/3+\beta_1'),$$

$$S_{3'n} = -S_1 e_z \cos(\varphi(n)+\pi/3+\alpha_1'), \qquad S_{6n} = -S_1(e_x+e_y)\cos(\varphi(n)+\pi/3+\beta_1').$$

Mixing coefficients for $\tau_4 \oplus \tau_6$ in case of +**k** from $\tau_4$:

a) $c_1 = 0.5 S_1(m_{1a},0;0,0;0,-am_{1a}^*;0,-am_{1a}^*)$, $c_2 = 0.5 S_1(-a^* n_{1a},0;0,0;0,0;0,-n_{1a}^*)$,

b) $c_1 = 0.5 S_1(0,0;m_{1b},0;0,-am_{1b}^*;0,-am_{1b}^*)$, $c_2 = 0.5 S_1(0,0;-a^* n_{1b},0;0,n_{1b}^*;0,n_{1b}^*)$,

(6k46c) = (6k46a)+(6k46b).

Mixing coefficients for $\tau_4 \oplus \tau_6$ in case of +**k** from $\tau_6$:

a) $c_1 = 0.5 S_1(0,m_{1a}'^*;0,m_{1a}'^*;m_{1a}',0;0,0)$, $c_2 = 0.5 S_1(0,0;0,-n_{1a}'^*;-n_{1a}',0;0,0)$,

b) $c_1 = 0.5S_1(0,-m_{1b}^*;0,0;0,0;m_{1b}',0)$, $c_2 = 0.5S_1(0,n_{1b}'^*;0,n_{1b}^*;0,0;-n_{1b}',0)$,

(6k46c') = (6k46a')+(6k46b').

2) Sites (6j):

For $\tau_2$ the basic results are quite similar to those of sites (6k):

$$\mathbf{S}_{1n} = S_2\mathbf{e}_x\cos(\varphi(n)+\alpha_{2a}), \qquad \mathbf{S}_{4n} = S_2\mathbf{e}_x\cos(\varphi(n)+\beta_{2a}),$$

(6j2a) $\quad \mathbf{S}_{2'n} = S_2\mathbf{e}_y\cos(\varphi(n)+\alpha_{2a}), \qquad \mathbf{S}_{5'n} = S_2\mathbf{e}_y\cos(\varphi(n)+\beta_{2a}),$

$$\mathbf{S}_{3'n} = -S_2(\mathbf{e}_x+\mathbf{e}_y)\cos(\varphi(n)+\alpha_{2a}), \qquad \mathbf{S}_{6'n} = -S_2(\mathbf{e}_x+\mathbf{e}_y)\cos(\varphi(n)+\beta_{2a}),$$

$$\mathbf{S}_{1n} = S_2\mathbf{e}_y\cos(\varphi(n)+\alpha_{2b}), \qquad \mathbf{S}_{4n} = S_2\mathbf{e}_y\cos(\varphi(n)+\beta_{2b}),$$

(6j2b) $\quad \mathbf{S}_{2'n} = -S_2(\mathbf{e}_x+\mathbf{e}_y)\cos(\varphi(n)+\alpha_{2b}), \qquad \mathbf{S}_{5'n} = -S_2(\mathbf{e}_x+\mathbf{e}_y)\cos(\varphi(n)+\beta_{2b}),$

$$\mathbf{S}_{3'n} = S_2\mathbf{e}_x\cos(\varphi(n)+\alpha_{2b}), \qquad \mathbf{S}_{6'n} = S_2\mathbf{e}_x\cos(\varphi(n)+\beta_{2b}),$$

$$\mathbf{S}_{1n} = S_2(\mathbf{e}_x+\mathbf{e}_y)\cos(\varphi(n)+\alpha_{2c}), \qquad \mathbf{S}_{4n} = S_2(\mathbf{e}_x+\mathbf{e}_y)\cos(\varphi(n)+\beta_{2c}),$$

(6j2c) $\quad \mathbf{S}_{2'n} = -S_2\mathbf{e}_x\cos(\varphi(n)+\alpha_{2c}), \qquad \mathbf{S}_{5'n} = -S_2\mathbf{e}_x\cos(\varphi(n)+\beta_{2c}),$

$$\mathbf{S}_{3'n} = -S_2\mathbf{e}_y\cos(\varphi(n)+\alpha_{2c}), \qquad \mathbf{S}_{6'n} = -S_2\mathbf{e}_y\cos(\varphi(n)+\beta_{2c}).$$

In case of $\tau_3\oplus\tau_5$ one obtains with $+\mathbf{k}$ from $\tau_3$:

$$\mathbf{S}_{1n} = S_2\mathbf{e}_z\cos(\varphi(n)+\alpha_2), \qquad \mathbf{S}_{4n} = S_2\mathbf{e}_z\cos(\varphi(n)+\beta_2),$$

(6j35) $\quad \mathbf{S}_{2'n} = -S_2\mathbf{e}_z\cos(\varphi(n)+\pi/3+\alpha_2), \qquad \mathbf{S}_{5'n} = -S_2\mathbf{e}_z\cos(\varphi(n)+\pi/3+\beta_2),$

$$\mathbf{S}_{3'n} = -S_2\mathbf{e}_z\cos(\varphi(n)-\pi/3+\alpha_2), \qquad \mathbf{S}_{6'n} = -S_2\mathbf{e}_z\cos(\varphi(n)-\pi/3+\beta_2) \text{ and}$$

with $+\mathbf{k}$ from $\tau_5$:

$S_{1n} = S_2 e_z \cos(\varphi(n)+\alpha_2')$, $\quad\quad S_{4n} = S_2 e_z \cos(\varphi(n)+\beta_2')$,

(6j35') $S_{2'n} = -S_2 e_z \cos(\varphi(n)-\pi/3+\alpha_2')$, $\quad S_{5'n} = -S_2 e_z \cos(\varphi(n)-\pi/3+\beta_2')$,

$S_{3'n} = -S_2 e_z \cos(\varphi(n)+\pi/3+\alpha_2')$, $\quad S_{6'n} = -S_2 e_z \cos(\varphi(n)+\pi/3+\beta_2')$.

Mixing coefficients for $\tau_4 \oplus \tau_6$ in case of $+\mathbf{k}$ from $\tau_4$:

a) $c_1 = 0.5 S_1(m_{1a},0;0,0;0,-a^* m_{1a}^*;0,-a^* m_{1a}^*)$, $c_2 = 0.5 S_1(n_{1a},0;0,0;0,0;0,-n_{1a}^*)$,

b) $c_1 = 0.5 S_1(0,0;m_{1b},0;0,-am_{1b}^*;0,-am_{1b}^*)$, $c_2 = 0.5 S_1(0,0;n_{1b},0;0,n_{1b}^*;0,n_{1b}^*)$,

(6j46c) = (6j46a)+(6j46b).

Mixing coefficients for $\tau_4 \oplus \tau_6$ in case of $+\mathbf{k}$ from $\tau_6$:

a) $c_1 = 0.5 S_1(0,-am_{1a}'^*;0,-am_{1a}'^*;m_{1a}',0;0,0)$, $c_2 = 0.5 S_1(0,0;0,-n_{1a}'^*;n_{1a}',0;0,0)$,

b) $c_1 = 0.5 S_1(0,am_{1b}'^*;0,0;0,0;m_{1b}',0)$, $c_2 = 0.5 S_1(0,n_{1b}'^*;0,n_{1b}'^*;0,0;n_{1b}',0)$,

(6j46c') = (6j46a')+(6j46b').

3) Sites (2e):

For $\tau_1$ and $\tau_2$: $c = 0.5 S_3(m_3, m_3^*)$, $m_3 = e^{i\alpha_3}$.

For $\tau_2$ results:

(2e2) $\quad S_{1n} = S_3 e_z \cos(\varphi(n)+\alpha_3) = -S_{2n}$.

In case of both $\tau_3 \oplus \tau_5$ and $\tau_4 \oplus \tau_6$:

a) $c = 0.5 S_3(m_3, a^* m_3^*; m_3, a m_3^*)$ and

b) $c = S_3(m_3, 0; 0, a m_3^*)$.

For $\tau_3 \oplus \tau_5$ one obtains with mixing coefficients $c = (S_3/2)(m_3,0;0,am_3^*)$, i.e. with $+\mathbf{k}$ from $\tau_3$:

(2e35) $\quad S_{1n} = S_3[e_x \cos(\varphi(n)+\alpha_3)+(1/\sqrt{3})(e_x+2e_y)\sin(\varphi(n)+\alpha_3)] = -S_{2n}$

or with c' = $(S_3/2)(m_3, a^*m_3^*; m_3, am_3^*)$, i.e. for +**k** from both $\tau_3$ and $\tau_5$:

(2e35') $\mathbf{S}_{1n} = S_3 \mathbf{e}_x \cos(\varphi(n)+\alpha_3) = -\mathbf{S}_{2n}$.

TABLE Ia. Refined lattice parameters a = 12.590(1) Å, c = 9.267(1) Å and atomic positional parameters of paramagnetic $Tb_{14}Ag_{51}$ at 30 K [7] for space group no. 175, *P6/m*. Estimated standard deviations of parameters are given within parentheses, refering to the last relevant digit. * Nominal reduced occupation according to ref. . [4]

| Atom | Site | x | y | z | Occup. |
|------|------|-----------|-----------|-----------|--------|
| Tb1  | 6k   | 0.1433(2) | 0.4703(2) | 0.5       | 100    |
| Tb2  | 6j   | 0.3906(3) | 0.1138(3) | 0         | 100    |
| Tb3  | 2e   | 0         | 0         | 0.3104(5) | 100    |
| Ag1  | 12l  | 0.0758(2) | 0.2663(2) | 0.2361(3) | 100    |
| Ag2  | 12l  | 0.1166(2) | 0.4943(2) | 0.1522(3) | 100    |
| Ag3  | 12l  | 0.4394(2) | 0.1034(2) | 0.3309(3) | 100    |
| Ag4  | 6k   | 0.2378(3) | 0.0588(4) | 0.5       | 100    |
| Ag5  | 6j   | -0.0018(6)| 0.1225(7) | 0         | 50*    |
| Ag6  | 4h   | 1/3       | 2/3       | 0.2994(5) | 100    |
| Ag7  | 2c   | 1/3       | 2/3       | 0         | 100    |

TABLE II. Irreducible representations and magnetic modes $\psi$ [9] for space group *P6/m*, no. 175, **k** = ±(1/3,1/3,0) and sites (6k), (6j) and (2e). a = $e^{i\pi/3}$, b = $e^{i\pi/6}$. Within parentheses the mode components along the basic translations of the chemical unit cell are given. With respect to magnetic neutron intensity calculations, it is for the present hexagonal symmetry in case of sites (6k) and (6j) more convenient to use instead of the magnetic atoms in the chemical unit cell (atoms designated without primes, see also Fig. 1) atom positions centered around the origin (marked by '), making use of equation (1).

| Sym. op./ Irred. rep. | 1 | $3^+$ | $3^-$ | $\bar{6}^+$ | m | $\bar{6}^-$ |
|---|---|---|---|---|---|---|
| $\tau_1$ | 1 | 1 | 1 | 1 | 1 | 1 |
| $\tau_2$ | 1 | 1 | 1 | -1 | -1 | -1 |
| $\tau_3$ | 1 | -a* | -a | 1 | -a* | -a |
| $\tau_4$ | 1 | -a* | -a | -1 | a* | a |
| $\tau_5$ | 1 | -a | -a* | 1 | -a | -a* |
| $\tau_6$ | 1 | -a | -a* | -1 | a | a* |

(6k):

2 = 2'+(1,1,0), 3 = 3'+(0,1,0), 4 = 4'+(1,0,0), 5 = 5'+(1,1,0)

| Atom | 1: | 2': | 3': | 4': | 5': | 6: |
|---|---|---|---|---|---|---|
| Position | x,y,z | -y,x-y,z | -x+y,-x,z | x-y,x,z | -x,-y,z | y,-x+y,z |
| Orbit | 1 | 1 | 1 | 2 | 2 | 2 |
| $\psi_{6k}(\tau_1,\mathbf{k})$ | (0,0,1) | (0,0,1) | (0,0,1) | (0,0,-a) | (0,0,-a) | (0,0,-a) |
| $\psi_{6k}(\tau_1,-\mathbf{k})$ | (0,0,-a) | (0,0,-a) | (0,0,-a) | (0,0,1) | (0,0,1) | (0,0,1) |
| $\psi_{a,6k}(\tau_2,\mathbf{k})$ | (1,0,0) | (0,1,0) | (-1,-1,0) | (-a,0,0) | (0,-a,0) | (a,a,0) |
| $\psi_{a,6k}(\tau_2,-\mathbf{k})$ | (0,a,0) | (-a,-a,0) | (a,0,0) | (1,1,0) | (-1,0,0) | (0,-1,0) |
| $\psi_{b,6k}(\tau_2,\mathbf{k})$ | (0,1,0) | (-1,-1,0) | (1,0,0) | (0,-a,0) | (a,a,0) | (-a,0,0) |
| $\psi_{b,6k}(\tau_2,-\mathbf{k})$ | (-a,-a,0) | (a,0,0) | (0,a,0) | (-1,0,0) | (0,-1,0) | (1,1,0) |
| $\psi_{6k}(\tau_3,\mathbf{k})$ | (0,0,1) | (0,0,-a) | (0,0,-a*) | (0,0,-a) | (0,0,-a*) | (0,0,1) |
| $\psi_{6k}(\tau_3,-\mathbf{k})$ | (0,0,1) | (0,0,-a) | (0,0,-a*) | (0,0,1) | (0,0,-a) | (0,0,-a*) |
| $\psi_{a,6k}(\tau_4,\mathbf{k})$ | (1,0,0) | (0,-a,0) | (a*,a*,0) | (-a,0,0) | (0,-a*,0) | (-1,-1,0) |
| $\psi_{a,6k}(\tau_4,-\mathbf{k})$ | (0,-1,0) | (-a,-a,0) | (a*,0,0) | (1,1,0) | (a,0,0) | (0,a*,0) |
| $\psi_{b,6k}(\tau_4,\mathbf{k})$ | (0,1,0) | (a,a,0) | (-a*,0,0) | (0,-a,0) | (a*,a*,0) | (1,0,0) |
| $\psi_{b,6k}(\tau_4,-\mathbf{k})$ | (1,1,0) | (a,0,0) | (0,a*,0) | (-1,0,0) | (0,a,0) | (-a*,-a*,0) |
| $\psi_{6k}(\tau_5,\mathbf{k})$ | (0,0,1) | (0,0,-a*) | (0,0,-a) | (0,0,-a) | (0,0,1) | (0,0,-a*) |
| $\psi_{6k}(\tau_5,-\mathbf{k})$ | (0,0,-a*) | (0,0,-a) | (0,0,1) | (0,0,1) | (0,0,-a*) | (0,0,-a) |
| $\psi_{a,6k}(\tau_6,\mathbf{k})$ | (1,0,0) | (0,-a*,0) | (a,a,0) | (-a,0,0) | (0,1,0) | (a*,a*,0) |
| $\psi_{a,6k}(\tau_6,-\mathbf{k})$ | (0,a*,0) | (-a,-a,0) | (-1,0,0) | (1,1,0) | (a*,0,0) | (0,a,0) |
| $\psi_{b,6k}(\tau_6,\mathbf{k})$ | (0,1,0) | (a*,a*,0) | (-a,0,0) | (0,-a,0) | (-1,-1,0) | (-a*,0,0) |
| $\psi_{b,6k}(\tau_6,-\mathbf{k})$ | (-a*,-a*,0) | (a,0,0) | (0,-1,0) | (-1,0,0) | (0,a*,0) | (-a,-a,0) |

(6j):

2 = 2'+(1,0,0), 3 = 3'+(1,1,0), 5 = 5'+(1,1,0), 6 = 6'+(0,1,0)

| Atom | 1: | 2': | 3': | 4: | 5': | 6': |
|---|---|---|---|---|---|---|
| Position | x,y,z | -y,x-y,z | -x+y,-x,z | x-y,x,z | -x,-y,z | y,-x+y,z |
| Orbit | 1 | 1 | 1 | 2 | 2 | 2 |
| $\psi_{6j}(\tau_1,\mathbf{k})$ | (0,0,1) | (0,0,1) | (0,0,1) | (0,0,1) | (0,0,1) | (0,0,1) |
| $\psi_{6j}(\tau_1,-\mathbf{k})$ | (0,0,1) | (0,0,1) | (0,0,1) | (0,0,1) | (0,0,1) | (0,0,1) |
| $\psi_{a,6j}(\tau_2,\mathbf{k})$ | (1,0,0) | (0,1,0) | (-1,-1,0) | (1,0,0) | (0,1,0) | (-1,-1,0) |
| $\psi_{a,6j}(\tau_2,-\mathbf{k})$ | (0,-1,0) | (1,1,0) | (-1,0,0) | (1,1,0) | (-1,0,0) | (0,-1,0) |
| $\psi_{b,6j}(\tau_2,\mathbf{k})$ | (0,1,0) | (-1,-1,0) | (1,0,0) | (0,1,0) | (-1,-1,0) | (1,0,0) |
| $\psi_{b,6j}(\tau_2,-\mathbf{k})$ | (1,1,0) | (-1,0,0) | (0,-1,0) | (-1,0,0) | (0,-1,0) | (1,1,0) |
| $\psi_{6j}(\tau_3,\mathbf{k})$ | (0,0,1) | (0,0,-a) | (0,0,-a*) | (0,0,1) | (0,0,-a) | (0,0,-a*) |
| $\psi_{6j}(\tau_3,-\mathbf{k})$ | (0,0,-a*) | (0,0,1) | (0,0,-a) | (0,0,1) | (0,0,-a) | (0,0,-a*) |
| $\psi_{a,6j}(\tau_4,\mathbf{k})$ | (1,0,0) | (0,-a,0) | (a*,a*,0) | (1,0,0) | (0,-a,0) | (a*,a*,0) |
| $\psi_{a,6j}(\tau_4,-\mathbf{k})$ | (0,a*,0) | (1,1,0) | (a,0,0) | (1,1,0) | (a,0,0) | (0,a*,0) |
| $\psi_{b,6j}(\tau_4,\mathbf{k})$ | (0,1,0) | (a,a,0) | (-a*,0,0) | (0,1,0) | (a,a,0) | (-a*,0,0) |
| $\psi_{b,6j}(\tau_4,-\mathbf{k})$ | (-a*,-a*,0) | (-1,0,0) | (0,a,0) | (-1,0,0) | (0,a,0) | (-a*,-a*,0) |
| $\psi_{6j}(\tau_5,\mathbf{k})$ | (0,0,1) | (0,0,-a*) | (0,0,-a) | (0,0,1) | (0,0,-a*) | (0,0,-a) |
| $\psi_{6j}(\tau_5,-\mathbf{k})$ | (0,0,-a) | (0,0,1) | (0,0,-a*) | (0,0,1) | (0,0,-a*) | (0,0,-a) |
| $\psi_{a,6j}(\tau_6,\mathbf{k})$ | (1,0,0) | (0,-a*,0) | (a,a,0) | (1,0,0) | (0,-a*,0) | (a,a,0) |
| $\psi_{a,6j}(\tau_6,-\mathbf{k})$ | (0,a,0) | (1,1,0) | (a*,0,0) | (1,1,0) | (a*,0,0) | (0,a,0) |
| $\psi_{b,6j}(\tau_6,\mathbf{k})$ | (0,1,0) | (a*,a*,0) | (-a,0,0) | (0,1,0) | (a*,a*,0) | (-a,0,0) |
| $\psi_{b,6j}(\tau_6,-\mathbf{k})$ | (-a,-a,0) | (-1,0,0) | (0,a*,0) | (-1,0,0) | (0,a*,0) | (-a,-a,0) |

(2e):

| Atom | 1: | 2: |
|---|---|---|
| Position | 0,0,z | 0,0,1-z |
| Orbit | 1 | 1 |
| $\psi_{2e}(\tau_1,\mathbf{k})$ | (0,0,1) | (0,0,1) |
| $\psi_{2e}(\tau_1,-\mathbf{k})$ | (0,0,1) | (0,0,1) |
| $\psi_{2e}(\tau_2,\mathbf{k})$ | (0,0,1) | (0,0,-1) |
| $\psi_{2e}(\tau_2,-\mathbf{k})$ | (0,0,1) | (0,0,-1) |
| $\psi_{2e}(\tau_3,\mathbf{k})$ | (b*,-i,0)/$\sqrt{3}$ | -(b*,-i,0)/$\sqrt{3}$ |
| $\psi_{2e}(\tau_3,-\mathbf{k})$ | (b,b*,0)/$\sqrt{3}$ | -(b,b*,0)/$\sqrt{3}$ |
| $\psi_{2e}(\tau_4,\mathbf{k})$ | (b*,-i,0)/$\sqrt{3}$ | (b*,-i,0)/$\sqrt{3}$ |
| $\psi_{2e}(\tau_4,-\mathbf{k})$ | (b,b*,0)/$\sqrt{3}$ | (b,b*,0)/$\sqrt{3}$ |
| $\psi_{2e}(\tau_5,\mathbf{k})$ | (b,i,0)/$\sqrt{3}$ | -(b,i,0)/$\sqrt{3}$ |
| $\psi_{2e}(\tau_5,-\mathbf{k})$ | (b*,b,0)/$\sqrt{3}$ | -(b*,b,0)/$\sqrt{3}$ |
| $\psi_{2e}(\tau_6,\mathbf{k})$ | (b,i,0)/$\sqrt{3}$ | (b,i,0)/$\sqrt{3}$ |
| $\psi_{2e}(\tau_6,-\mathbf{k})$ | (b*,b,0)/$\sqrt{3}$ | (b*,b,0)/$\sqrt{3}$ |

**Figure Captions**

FIG. 1. a) Tb sublattices of $Tb_{14}Ag_{51}$ at 30 K with atom numbers according to Table I. The other numbers indicate characteristic interatomic distances, which imply considerable frustrations in magnetic interactions. Longer distances are shown as thinner lines. A chemical unit cell is outlined.

b) A view perpendicular to the c-axis corresponding to the magnetic unit cell, with hexagonal Tb2 (inner, at z = 0) and Tb1 (outer, at z = 1/2) rings. Tb3 is shown with smaller diameter. With thick lines also the chemical unit cell is indicated. The unit cell based on the diagonal of the chemical cell represents a hexagonal H cell as possible smallest magnetic unit cell.

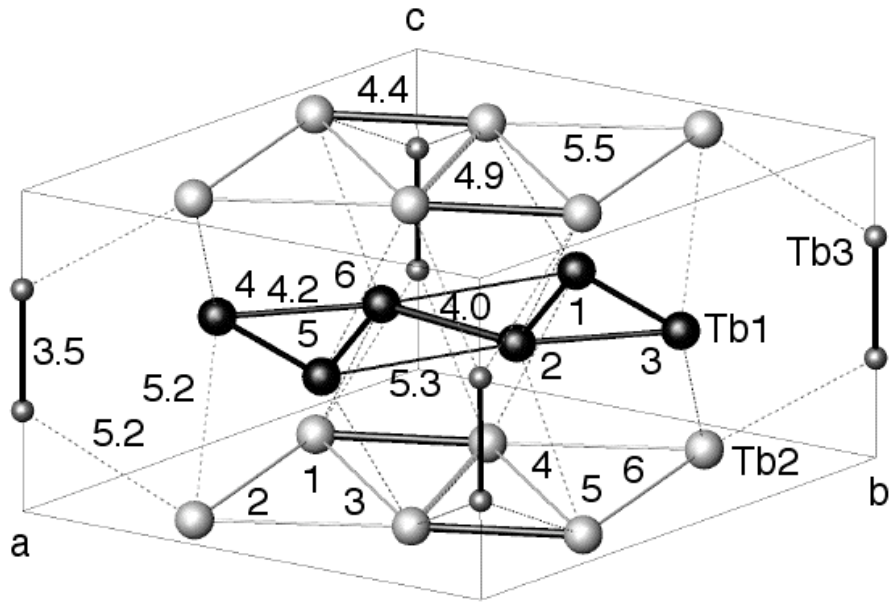

Fig. 1a

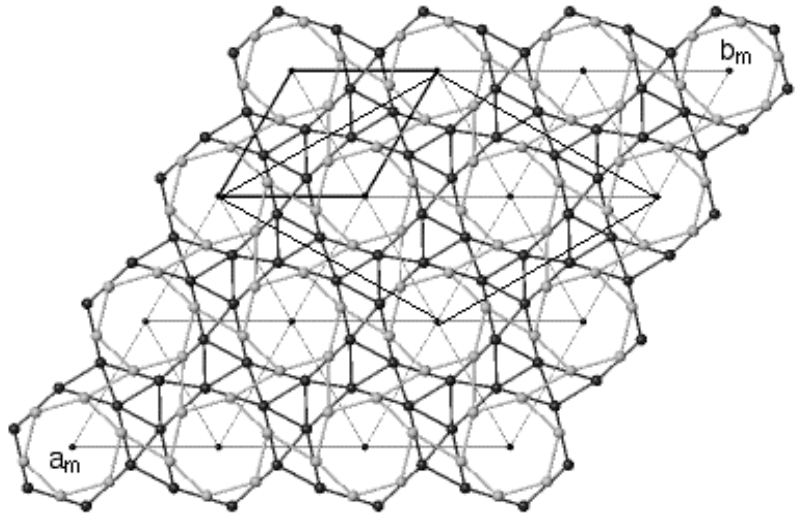

Fig. 1b